# Non-invariance of the speed of light in free-space

N.I. Petrov

*Scientific and Technological Center of Unique Instrumentation of Russian Academy of Sciences, Moscow 117342, Russia*
*petrovni@mail.ru*

**Abstract:** A plane monochromatic wave propagates in vacuum at the velocity of *c*. However, wave packets limited in space and time are used to transmit energy and information. Here it has been shown based on the wave approach that the on-axis part of the pulsed beams propagates in free space at a variable speed, exhibiting both subluminal and superluminal behaviours in the region close to the source, and their velocity approaches the value of *c* with distance. Although the pulse can travel over small distances faster than the speed of light in vacuum, the average on-axis velocity, which is estimated by the arrival time of the pulse at distances $z >> l_d$ ( $l_d$ is the Rayleigh diffraction range) and $z > c\tau$ ( $\tau$ is the pulse width) is less than *c*. The total pulse beam propagates at a constant subluminal velocity through the whole distance. The mutual influence of the spatial distribution of radiation and temporal shape of the pulse during nonparaxial propagation in vacuum is studied. It is found that the decrease in the width of the incident beam and the increase in the central wavelength of the pulse lead to a decrease in the propagation velocity of the wave packet. It is shown that the velocities of the Bessel-Gauss and Laguerre-Gauss pulsed beams decrease with increasing orbital angular momentum (OAM). The results obtained are in good agreement with subluminal and superluminal measurements in free space.

## 1. Introduction

The pulse wave packet propagation effects are of considerable interest because of their significance from fundamental as well as application aspects such as optical communication, information science, digital holography, image processing, etc.

Spatially and temporally localized beams are of interest since they describe the fields associated with dielectric antennae, laser optical systems and other sources emitting focused radiation more realistically than plane waves. It is known that the spatiotemporal shape of such wave-packets is being modified during the propagation due to the effects of non-stationary diffraction. One of the reasons leading to the transformation of the space-time radiation structure is the interconnected diffraction transformation of the frequency and angular spectra. Many works are devoted to the study of the propagation of radiation pulses in various media [1-5]. Non-stationary diffraction effects arising in the propagation of electromagnetic pulses in free space are considered in [2]. In [3-4], the problem of diffraction of a pulsed Gaussian beam in vacuum is solved in the paraxial approximation, and the relationship of spatial and temporal characteristics of the beam is investigated on the basis on the spectral approach. In [5] an approach to the description of nonstationary diffraction of extremely short pulses based on a generalization of the stationary Sommerfeld diffraction theory is developed.

In recent years, the study of superluminal and subluminal propagation of ultrashort pulses in free space is of particular interest. The speed of light propagation in vacuum is one of the fundamental characteristics of electromagnetic waves. A plane monochromatic wave propagates in vacuum at the velocity *c*. However, wave packets limited in space and time are used to transmit energy and information. The wave packet propagates at a group velocity that is different from the velocity of individual harmonic components. Phase and group velocities of light pulses can differ significantly in dispersive media such as cold atomic clouds [6], atomic vapors [7-9], etc. In [10] it was demonstrated experimentally and theoretically the possibility of superluminal propagation of the pulse maximum in an amplifying medium. The superluminal effects were also observed in vacuum. In [11] the superluminal group velocity of an ultrashort (70 fs) optical Bessel beam pulse was measured. In [12] the superluminal phase and group velocities of the Bessel pulse beam were determined using an interferometric setup.

The effect of the transverse spatial structure of the light beam on its propagation velocity was experimentally discovered in [13]. The effect was explained by the delay of peripheral regions of the beam using the geometric optics approach. However, such an approach is insufficient, and a rigorous analysis of the problem is possible only within the framework of wave optics, taking into account nonstationary diffraction effects [14]. It was shown in [15] that the slowing down of light also depends on the magnitude of the orbital momentum of the beam. The group velocities manifested by Laguerre-Gauss (LG) modes in vacuum are investigated and the subluminal effects arising from the twisted nature of the optical phase front are observed and explained. However, LG functions are not the solutions of the Helmholtz wave equation in free space, and they can be considered as the modes in vacuum only in paraxial approximation.

In this paper, the theoretical analysis of the propagation of vortex pulsed beams in free space in the framework of the wave approach is carried out taking into account non-stationary effects of diffraction and nonparaxiality. The mutual influence of the spatial distribution of the incident radiation and the temporal shape of the pulse during vacuum propagation is studied. In particular, the subluminal effects arising from the spatial localization of the pulse beam are revealed. The smaller the beam radius, the slower the propagation speed. A strong change in the pulse shape as a consequence of nonstationary diffraction is demonstrated for tightly focused beams. The spatial modes with azimuthal and radial indices are proposed to use for simulation of the pulsed beam propagation in free space. Mode representation provides physical insight and computational simplification in the analysis of pulsed beams in free space. The influence of the spatial limitation of the pulse beam on the propagation velocity is studied. It is shown that there is a significant difference between the axial velocity (the velocity measured on the propagation axis at one point of the beam cross-section) and the total pulse beam velocity (when the receiver captures the full cross-section). On-axis velocity exhibits both superluminal and subluminal behaviors along the propagation distance, whereas the total cross-section velocity is subluminal through the whole distance.

The results of this work extend the known results of the propagation of pulsed beams in free space and can be applied in many areas of optics and photonics, such as optical communication, temporal imaging, supercontinuum generation, etc.

## 2. Basic equations

The Maxwell equations describing the propagation of light are reduced to

$$\nabla\times\nabla\times\vec{E}=-\frac{1}{c^2}\frac{\partial^2\vec{E}}{\partial t^2}\quad . \tag{1}$$

For a quasi-monochromatic field, we find that the Fourier components $E(\vec{r},\omega-\omega_0)$, which are defined as

$$E(\vec{r},\omega-\omega_0)=\int_{-\infty}^{\infty}E(\vec{r},t)\exp\left[i\left(\omega-\omega_0\right)t\right]dt\,, \tag{2}$$

satisfy the Helmholtz equation, where $\omega_0$ is the center frequency of the pulse.

First, we find the evolution of each spectral component of the spatiotemporal incident pulse. It is known that the diffraction-free Bessel beams are the solutions of the Helmholtz wave equation [16, 17]. They can be considered as modal solutions with azimuthal indices in free space. However, Bessel beams have infinite transverse size and require infinite power. In practice, quasi-Bessel beams of limited transverse dimensions generated by an axicon or conical lens are used. Such beams exhibit no diffraction over a limited propagation distance [16, 17]. There are also modal solutions of finite transverse size with discrete azimuthal and radial indices similar to modal solutions in cylindrical waveguides. The transverse field profiles of these solutions remain invariant along the effective depth of field. Note that these solutions form a complete set of mutually orthogonal functions in a given interval [0, $R_0$]. Hence, any field in the initial plane $z = 0$ can be decomposed into these modal solutions.

The normalized Bessel functions with radial $p$ and azimuthal $l$ indices can be considered as the modal solutions over the diffraction-free region

$$\psi_{pl}(\rho,\varphi) = J_l(\mu_{pl}\,\rho/R_0)\exp(il\varphi)\Big/\left(\sqrt{\pi}R_0 J_{l+1}(\mu_{pl})\right), \qquad (3)$$

where $\mu_1, \mu_2, ...$ are the positive zeros of the Bessel function $J_l(z)$.

The evolution of the electric field is determined by the expression

$$E(\rho, z, \omega-\omega_0) = F(\omega-\omega_0)\sum_{pl} c_{pl}\psi_{pl}(\rho)\exp(i\beta_{pl}(\omega)z), \qquad (4)$$

where $F(\omega-\omega_0)$ is the frequency spectrum of the incident pulse, $c_{pl}$ are the modal coefficients depending on the incident field parameters, $\beta_{pl}(\omega)$ are the propagation constants of the modes with radial $p$ and azimuthal $l$ indices, respectively.

Expand $\beta(\omega)$ in a Taylor series in the neighbourhood of $\omega_0$:

$$\beta_p(\omega) = \sum_{m=0} \frac{(\omega-\omega_0)^m}{m!}\gamma_{m,p}, \qquad (5)$$

where $\gamma_{m,p} = \frac{d^m}{d\omega^m}\beta_p(\omega)\Big|_{\omega=\omega_0}$, $\gamma_{1,p} = \frac{d\beta_p}{d\omega} = \frac{1}{c}\frac{d\beta_p}{dk}$.

Consider the incident pulse in the plane $z = 0$, whose envelope is described by a function $f(t) = \exp\left(-t^2/2\tau^2 + i\omega_0 t\right)$, where $\tau$ is the width of the input pulse.

The inverse Fourier transform gives an expression for the electric field in the time domain:

$$E(\rho, z, t) = \left(\frac{\tau}{\sqrt{2\pi}}\right)\sum_{p=1}^{N} c_{pl}\psi_{pl} f(t, z, \tau), \qquad (6)$$

где $f(t,z,\tau) = \frac{\exp\left[i\left(t-\gamma_{0,p}z\right)\right]}{\left(\tau^2 + i\gamma_{2,p}z\right)^{1/2}}\exp\left\{-\frac{\left(t-\gamma_{1,p}z\right)^2}{2\left(\tau^2 + i\gamma_{2,p}z\right)}\right\}$.

The coefficients $c_{pl}$ are determined by the incident pulse field $E(\rho,0,t) = E(\rho,0)f(t)$:

$$c_{pl} = \int_0^{\infty}\int_0^{2\pi} E(\rho,0)\psi_{pl}^*(\rho,\varphi)\rho\, d\rho\, d\varphi. \qquad (7)$$

The integrals (7) for the pulsed beams with Gaussian, Bessel-Gauss and Laguerre-Gauss spatial distributions can be calculated analytically. A limited number of modes $N$ can be considered in summation (6).

It can be seen from (6) that the electric field of each mode reaches the maximum value at

$$t = z\gamma_1 = z\frac{d\beta}{d\omega} = \frac{z}{c}\frac{d\beta}{dk},$$

where $v_g = c\left(\frac{d\beta}{dk}\right)^{-1} = c\left[1 - \mu_{pl}^2/\left(R^2k^2\right)\right]^{1/2}$ is the group velocity of the modes.

The group velocity of the mode acquires a maximum value of *c*, when $\lambda \to 0$, which corresponds to the approximation of geometric optics.
The phase velocity of the mode is equal to:

$$v_{ph} = \frac{\omega}{\beta} = c\left[1 - \mu_{pl}^2/\left(R^2k^2\right)\right]^{-1/2}.$$

It is seen that $v_{ph} > c$ and $v_g v_{ph} = c^2$. The higher the mode number, the greater the phase velocity. However, this does not mean that energy or information can be transmitted at this rate.
The intensity distribution $I(\rho,z,t) = \left|E(\rho,z,t)\right|^2$ is determined by the expression

$$\begin{aligned} I(\rho,z,t) &= I_0 + I_1 = \frac{1}{2\pi}\sum_{p=1}^{N}\left|c_{pl}\right|^2\left|\psi_{pl}\right|^2\left|Q_{pl}\right|^2 \\ &+ \frac{1}{2\pi}\sum_{p=1}^{N}\sum_{n=1}^{N} c_{pl}^{*}c_{nl}\psi_{pl}^{*}\psi_{nl}Q_{pl}^{*}Q_{nl} \end{aligned}, \quad p \neq n, \qquad (8)$$

where $Q_{pl} = \frac{\exp\left[i\left(t-\gamma_{0,p}z\right)\right]}{\left(\tau^2 + i\gamma_{2,p}z\right)^{1/2}}\exp\left\{-\frac{\left(t-\gamma_{1,p}z\right)^2}{2\left(\tau^2 + i\gamma_{2,p}z\right)}\right\}$.

It is seen that the first term in (8) is the sum of the intensities of the modes and contains only the group velocities of the modes, while the second term, representing the interference of the modes, includes both the phase $v_{ph} = 1/\gamma_{0,p}$ and group $v_g = 1/\gamma_{1,p}$ velocities. This indicates that only the second term is responsible for the superluminal effect that occurs due to interference between modes.
If the integration of intensity across the entire cross-section is taken, then the power (total intensity) $P(z,t)$ is determined by

$$\left\langle I(\rho,z,t)\right\rangle = \frac{1}{2\pi}\Sigma\left(\left|c_{pl}\right|^2/b\right)\exp\left\{-\frac{\left(t-\gamma_{1,p}z\right)^2}{b^2\tau^2}\right\}, \qquad (9)$$

где $b = \left(1 + \gamma_{2,p}^2 z^2/\tau^4\right)^{1/2}$.

The cross-terms in (8) describing the interference between different modes become zero due to the orthogonality condition. Consequently, the mode intensity components in (9) reach their maximum values, when $t = z\gamma_{1,p} = z/v_g$, i.e. only the subluminal propagation can be observed by measuring the total intensity over the entire beam cross-section.
Wave propagation is characterized by various velocities: the phase, group, signal envelope amplitude, and energy. The velocities for pulse amplitude, pulse center of gravity, and pulse energy flow can be considered for pulse beams:

$v_m = z/t_m$, $v_{c.g.} = z/T_{c.g.}$ and $v_E = S/w$,

where $t_m$ is the arrival time of the pulse amplitude, $T_{c.g.}$ is the arrival time of the pulse center of gravity, *S* is the Poynting vector, and *w* is the electromagnetic energy density.

The pulse velocity can be determined from equation (8). It follows from this that both the group and phase velocities of the modes contribute to the resulting pulse velocity if only part of the beam cross-section is recorded. This means that the velocity of the pulse beam depends on the size of the receiver aperture.
The energy flow velocity depends only on the group velocities of the modes, so the averaged energy flow velocity $v_E$ of a pulsed beam is always subluminal, i.e. $v_E < c$. In addition, the instantaneous (local) velocity of propagation can be defined as $v_{ins} = dz/dt$ at different distances along the axis of propagation.
The arrival time can be determined both for a given beam cross-section point and for the entire beam, i.e. by averaging over the cross-section.
The arrival time of the pulse center of gravity, which is determined by $t_{c.g.} = \bar{t}(\rho, z) = \int I(\rho, z, t) t dt / \int I(\rho, z, t) dt$, is different for various points of the beam cross-section. The on-axis arrival time is equal to $T_{ar}^{axis} = \bar{t}(0, z)$. If the total cross-section of the pulse beam is recorded, then arrival time is determined by averaging over the entire cross-section:
$T_{ar}^{tot} = \langle t(z) \rangle = \iint I(\rho, z, t) \rho d\rho t dt / \iint I(\rho, z, t) \rho d\rho dt$.
The average time of arrival of the center of gravity of the total beam can be calculated analytically: $T_{ar}^{tot} = \langle t \rangle = \sum_{p=1}^{N} |c_{pl}|^2 \gamma_{1,p} z$. The velocity of the center of gravity of the total beam is given by $v_{c.g.}^{tot} = z / \langle t \rangle = \left( \sum_{p=1}^{N} |c_{pl}|^2 \gamma_{1,p} \right)^{-1}$, where $\gamma_{1,p}$ is determined by the group velocities of the modes, and hence $v_{c.g.}^{tot} < c$. It is seen that this velocity depends on the parameters of the incident pulse beam and remains constant during propagation.
The average width of pulses changes during propagation. The modal dispersion of the pulse beam, defined as $\sigma_\tau^2 = \left\langle (\Delta t)^2 \right\rangle = \langle t^2 \rangle - \langle t \rangle^2$, is given by

$$\sigma_\tau^2 = \sum_{p=1}^{N} |c_{pl}|^2 \left( \gamma_{1,p}^2 z^2 + \tau^2/2 + \gamma_{2,p}^2 z^2 / (2\tau^2) \right) - \left( \sum_{p=1}^{N} |c_{pl}|^2 \gamma_{1,p} z \right)^2 . \quad (10)$$

Thus, the pulse width increases with the propagation distance depending on the group velocities, the dispersion of the group velocities, and the pulse duration.

## 3. Simulation results

The spatiotemporal shape of wave packets is being modified during the propagation due to the nonstationary effects of diffraction and interference of modes. A strong change in the pulse shape occurs when the beam width and pulse duration decrease. Note that the width of the beam in front of the pulse varies less than in the tail part during the propagation. This indicates that the frontal part (high frequencies) of the pulses is less affected by diffraction.
Although the simulation can be performed using the integral representation, the simulation results using the modal approach are presented below. The mode representation provides a simplification of calculations in the analysis of pulsed beams in free space.
In Fig. 1 the pulse intensities at different distances $z_k$ are presented. In Fig.1b, c the BG pulses are presented in an offset time scale $t - (z_k - z_1)/c$, where $z_k$ is the distance, at which the pulse is recorded. It is seen that the pulses have the same shape at different distances, but they are offset relative to each other. These shifts originate from differences in the time of arrival of the pulse amplitude. This means that the considered pulsed beam experiences the subluminal propagation. As the width of the incident beam decreases, the time shift of the amplitude position between the considered pulse and plane-wave propagating at the velocity $c$ increases.

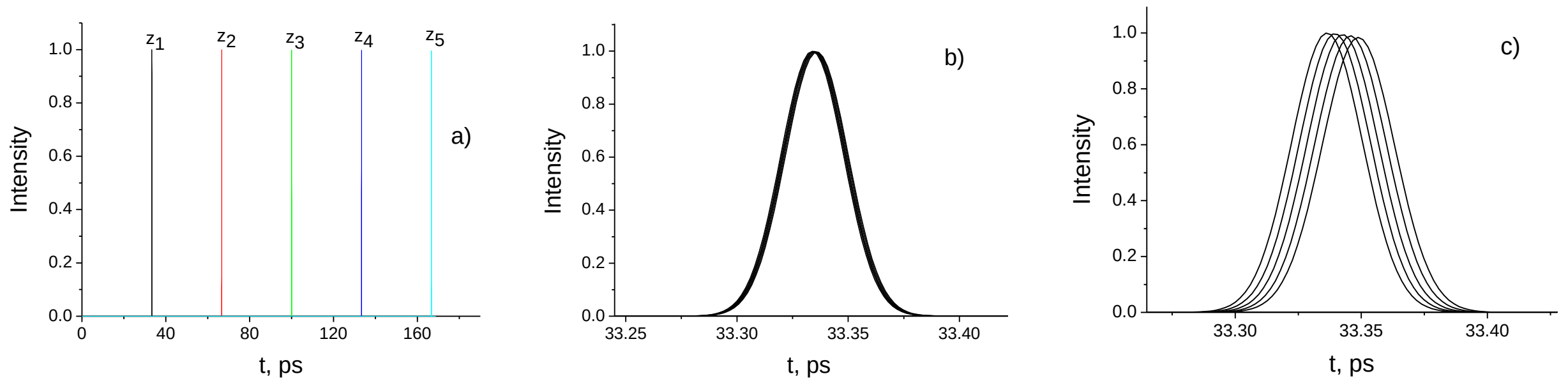

Fig. 1. (a) Pulse shapes at different distances $z_k$ = 10, 20, 30, 40, 50 mm. (b, c) Pulse intensities at the distances $z_k$ = 10, 20, 30, 40, 50 mm in the coordinate system with time delay $t-\left(z_k-z_1\right)/c$ . $\tau$ = 20 fs; (b) $a_0$ = 100 μm is the radius of Gaussian distribution; $w_B$ = 50 μm is the central peak width of the Bessel function; (c) $a_0$ = 100 μm; $w_B$ = 20 μm.

In Fig. 2 the delay times of the pulsed BG and LG beams (the arrival time of the pulse center of gravity compared to light in vacuum) $\Delta T = T_{ar}^{tot} - T_0$, where $T_0 = z/c$, depending on the distance $z$ are presented for different values of OAM. It is seen that the delays are higher for pulsed BG beams and increase with increasing OAM.

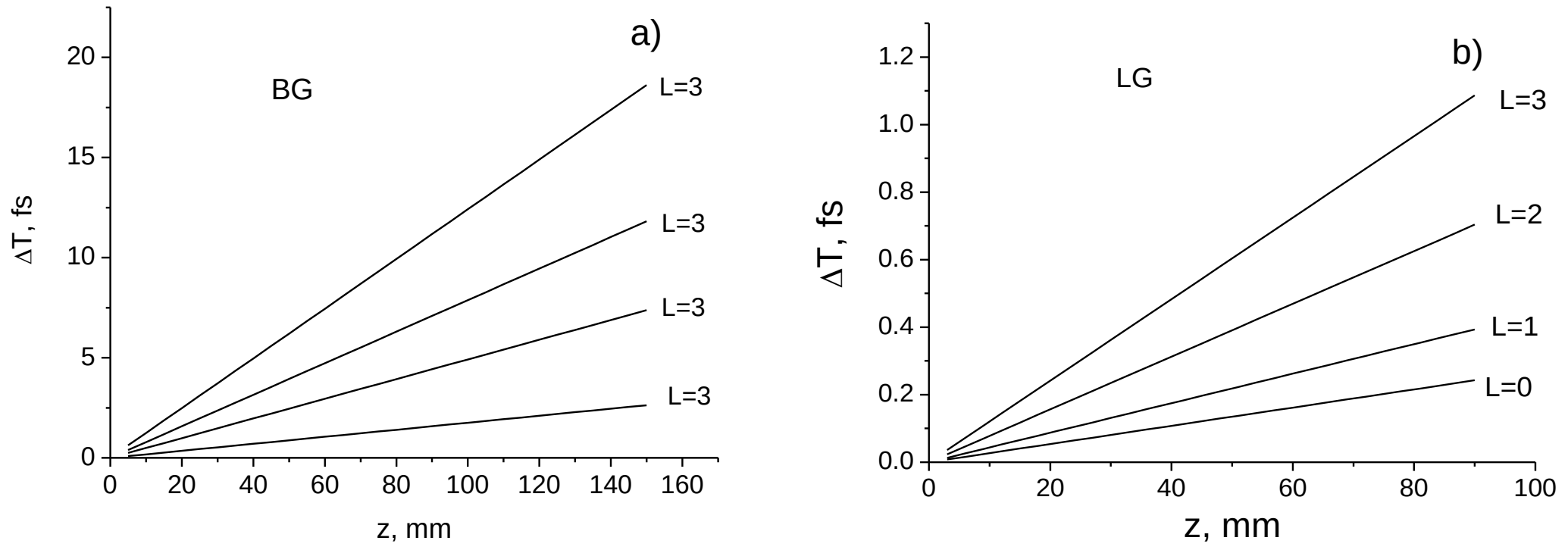

Fig. 2. Delays in arrival time as a function of distance $z$ for the pulsed BG (a) and LG (b) beams: $a_0$ = 100 μm; $w_B$ = 100 μm; $w_0$ = 100 μm; $\lambda$ = 795 nm.

In Fig. 3 the dependences of the arrival times of BG pulsed beams relative to a plane-wave pulse on the cone angle $\theta$ (a), the central peak width (beam radius) $w_B$, and orbital angular momentum $L$ at the propagation distance $z$ = 1 m are presented. The pulse duration $\tau$ = 100 fs and wavelength $\lambda$ = 710 nm.

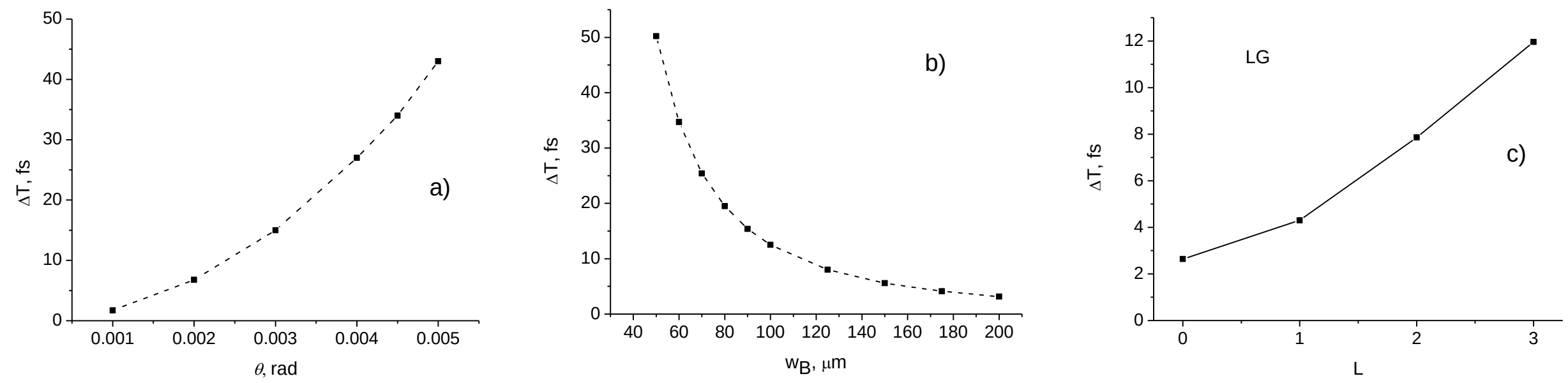

Fig. 3. Arrival time delays as a function of the cone angle $\theta$ (a), central peak width $w_B$ (b), and orbital angular momentum $L$ (c) relative to the plane-wave case. The beam parameters are: (a, b) pulsed BG beam with $\tau$ = 100 fs, $L$ = 0, $\lambda$ = 710 nm; (c) pulsed LG beam with $\tau$ = 100 fs, $w_0$ = 100 μm, $\lambda$ = 795 nm.

It follows from the calculations that a decrease in the velocity compared to the velocity of light $c$ leads to a delay of $\delta z_B \approx \frac{z}{2}\theta^2 \approx \frac{z}{2}\frac{\lambda^2}{w_B^2}$, where $\theta = \sin^{-1}(\alpha / k)$, $w_B \approx 2.4/(k\sin\theta)$ is the central peak width of a Bessel beam, and $k = 2\pi/\lambda$, $\lambda$ is the central wavelength. Consequently, for the time delay and the velocity reduction, we obtain $\delta T \approx (z/c)\lambda^2/w_B^2$ and $\delta v \approx c\,\lambda^2/w_B^2$.

These results are in good agreement with the experimental data [13, 15]. Time delays presented in Fig.6a agree well with the measured delays for the Bessel beam in [13]. In [13] it was reported that the group delay increases with the square of the diameter of the Gaussian beam. Here we have shown that the delay time increases as the diameter of the incident beam decreases. However, there is no contradicction between these results. The point is that the incident beam in [13] does not propagate in free space, but it is focused by the lens. The spot size of the focused beam that is responsible for the delay is inversely proportional to the waist of the input beam: $w_0 = f\lambda / w_{in}$, where $f$ is the focal length of the lens, $w_{in}$ is the waist of the input beam. Therefore, in the absence of a lens, the larger the diameter of the input beam, the smaller the waist of the focused beam. The calculated delay times depending on OAM are also in good agreement with the measurements [15], if the paraxial beams with $w_0 >> \lambda$ are considered.

In Fig. 4a the delay times $\delta T^{axis} = T_{ar}^{axis} - T_0$ and $\delta T^{tot} = T_{ar}^{tot} - T_0$, where $T_0 = z/c$, depending on the distance $z$ are presented for the pulsed BG beam. The delay of the center of gravity of the entire pulse beam relative to the plane wave increases linearly with the distance (Fig. 4a). However, the delay of the center of gravity of the axial part of the pulse beam increases only in the region close to the source, and it disappears with increasing propagation distance. In Fig. 4b the on-axis $v_{ar}^{axis}$, $v_{ins}^{axis}$, and total cross-section $v_{ar}^{tot}$ velocities are presented. The total cross-section velocity depends on the initial parameters of the pulsed beam and remains constant during propagation (Fig. 4b, c).

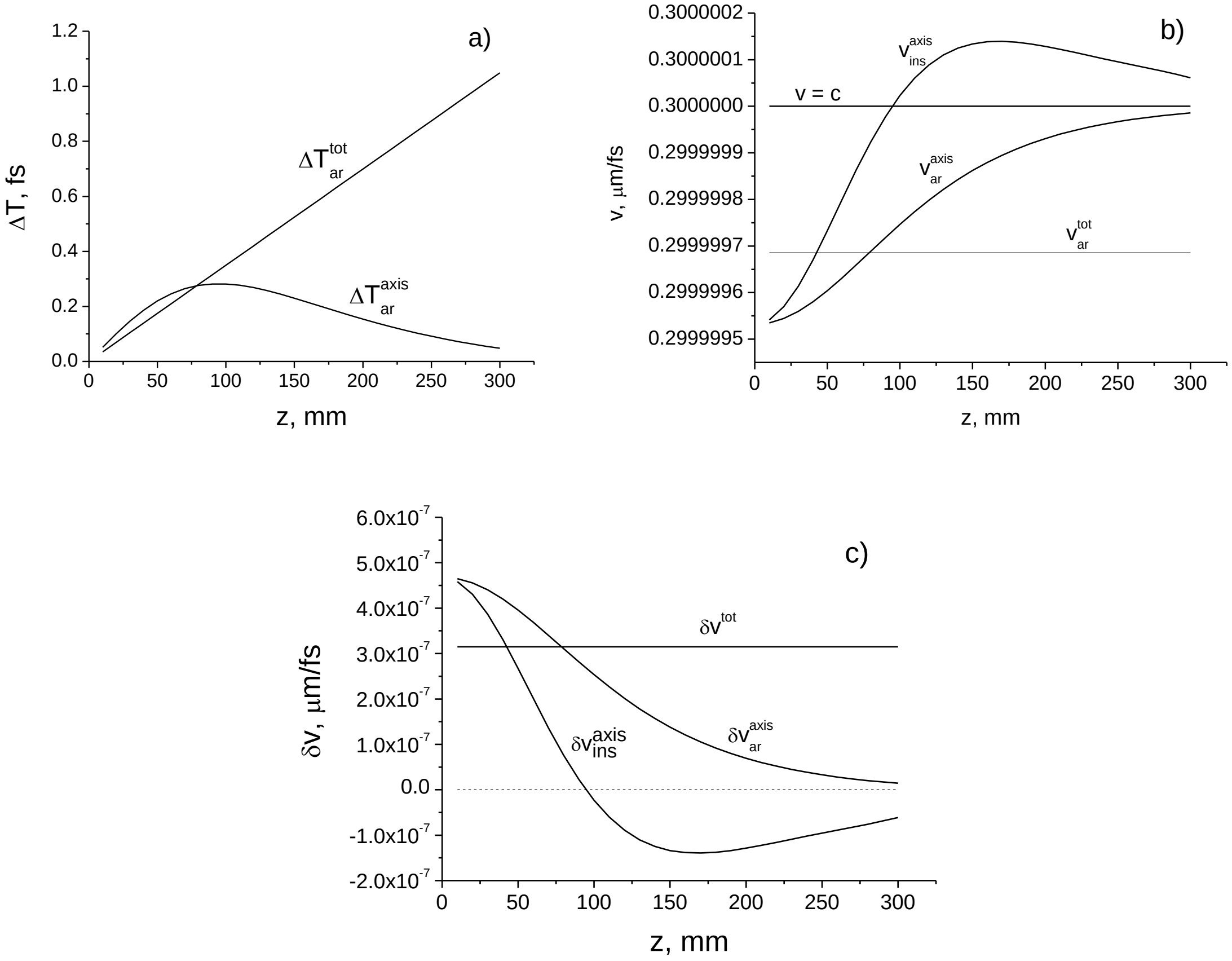

Fig. 4. The delays in arrival time (a), the velocities (b), and the velocity changes (c) as a function of the propagation distance *z*. The beam parameters are: $a_0$ = 200 μm; $w_B$ = 200 μm; $\lambda$ = 710 nm.

It is seen that the total beam cross-section velocity $v_{ar}^{tot}$ is less than *c* for spatially structured pulsed beams. However, the on-axis velocity determined on the optical axis in a single point varies with the distance from the source. In this case the instantaneous on-axis velocity $v_{ins}^{axis}$ can be higher than *c*, even though the velocity $v_{ar}^{axis}$ is subluminal (Fig. 4b). Indeed, it follows from the calculation that the instantaneous on-axis velocity of the pulse becomes superluminal in the Rayleigh diffraction region. The superluminality gradually disappears with distance due to the vanishing of the interference term in (8) (Fig. 4b, c). This indicates that superluminal propagation is due to interference between different modes (second term in (8)). For $z >> l_d$ the average velocity which is estimated as $\bar{v}_{ins}^{axis} = \left(1/d\right)\int_0^d v_{ins}^{axis}(z)dz$, where $T_{ar}$ is the time of arrival of the pulse center of gravity, is less than *c* (Fig. 4b). The velocity difference $\delta v_{ar}^{axis} = c - v_{ar}^{axis}$ decreases, approaching zero with increasing distance (Fig. 4c). The on-axis velocity of the center of gravity $v_{axis} = c - \delta v_{axis}$ increases with distance *d*, since the value $\delta v_{axis}$ decreases with distance (Fig. 4). When $z \to \infty$, the velocity difference $\delta v_{axis} \to 0$, so the on-axis velocity $v_{axis} \to c$. Note that the distance region, where the instantaneous on-axis velocity is higher than *c*, decreases with the decrease of the beam cross-section radius.

In Fig. 5 the delay times and velocities as a function of distance $z$ for pulse BG, LG and Gauss beams are presented. It is seen that the velocities $v_{ar}^{axis}$ and $v_{ar}^{tot}$ in the region close to the source are lower for the BG beam as compared to the LG and Gauss beams with the same beam waists.

It is seen that the on-axis velocity varies with the distance approaching the value of $c$ at $z > l_d = kw_0^2/2$, where $l_d$ is the Rayleigh diffraction length. The total cross-section velocity $v_{ar}^{tot}$ decreases with wavelength and remains constant throughout the propagation distance. The velocity $v_{ar}^{tot}$ decreases with the decrease of the beam waist and does not change with distance.

The spatial delays can be approximated by the expression $\delta z \approx z\lambda^2/w_0^2$. Consequently, for the time delay and the velocity reduction, we obtain $\delta T \approx (z/c)\lambda^2/w_0^2$ and $\delta v \approx c\,\lambda^2/w_0^2$. Note that the delays increase as the wavelength increases of and the beam width decreases. When $w_0 \to \infty$, $\delta v \to 0$, therefore there is no delay of the pulse relative to the plane-wave case. Similarly, $\delta v \to 0$, when $\lambda \to 0$. This indicates that short wavelengths propagate with the speed close to the plane wave velocity $c$, and long wavelengths undergo a delay.

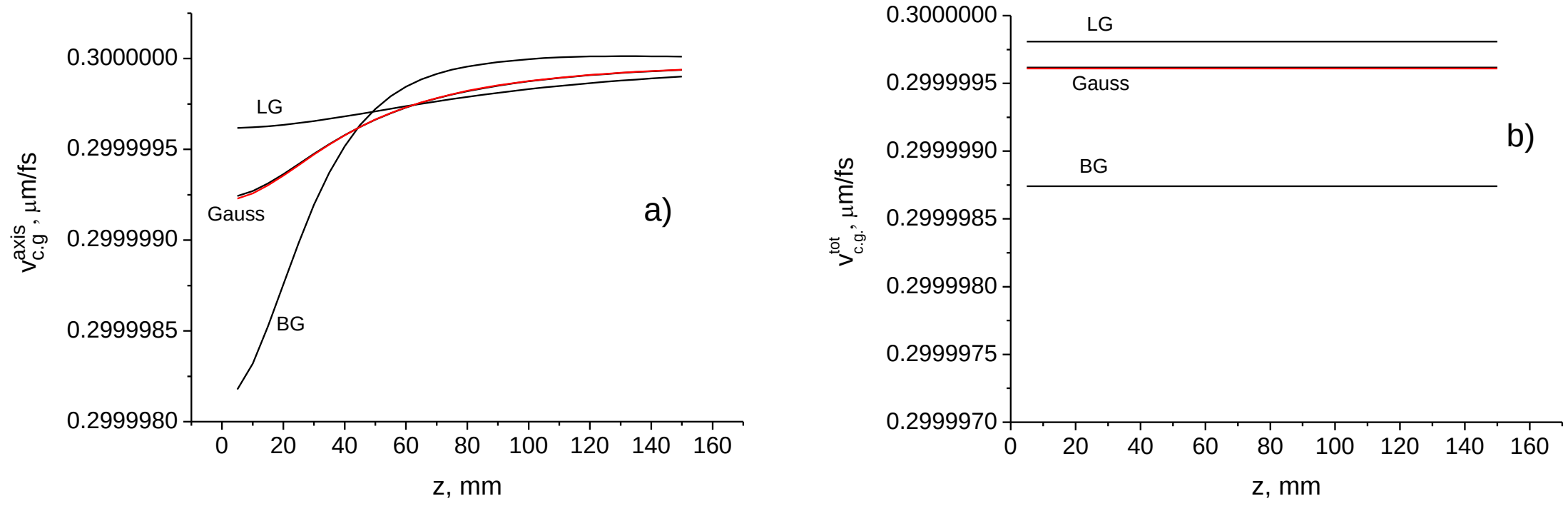

Fig. 5. On-axis (a) and total cross-section (b) velocities as a function of propagation distance for BG, LG, and Gauss pulsed beams. The beam parameters are: BG - $a_0$ = 100 μm; $w_B$ = 100 μm; LG - $w_0$ = 100 μm – black line, $w_0$ = 70 μm – red line; Gauss - $a_0$ = 100 μm. $\lambda$ = 710 nm.

In Fig. 6 the pulse shapes on the propagation axis ($\rho = 0$) corresponding to the terms $I_0$ and $I_1$ in (8) and their sum are presented for the BG pulse beam. It is seen that the pulse associated with the interference term $I_1$ propagates at a higher rate than the pulse corresponding to the term $I_0$. Therefore, in this case, superluminal behavior may be observed. This indicates that the local pulse velocity may be higher than $c$. However, the average velocity, which is estimated by the time of arrival of the pulse at $z > l_d$, is less than $c$.

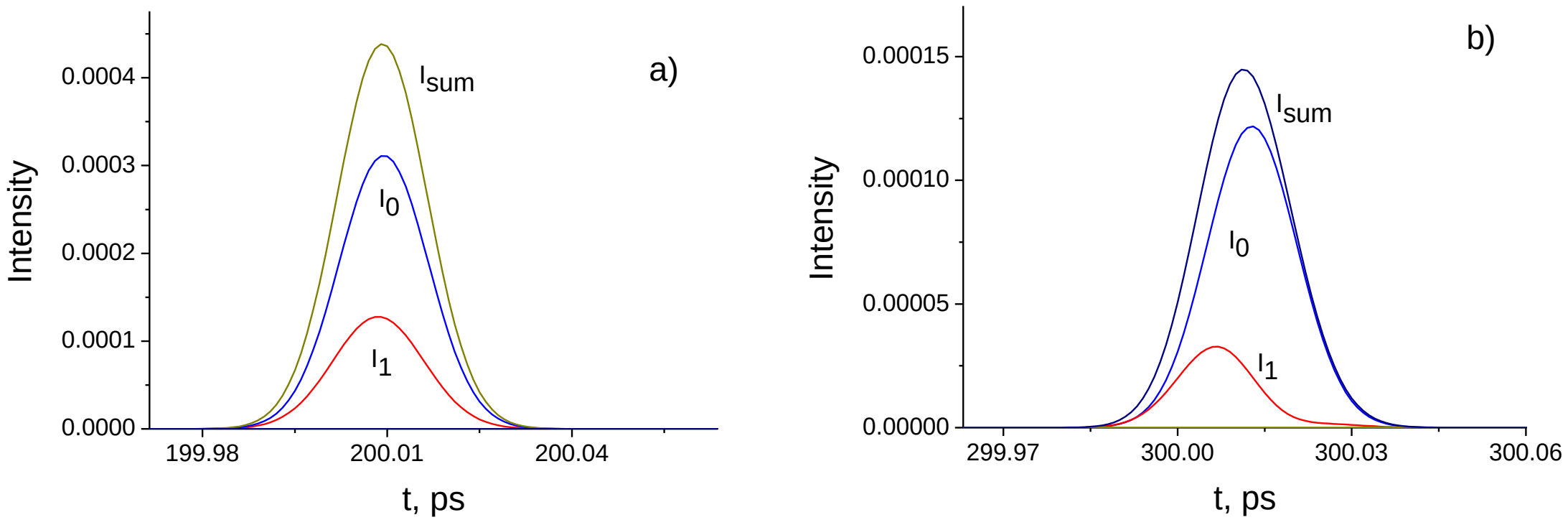

Fig. 6. On-axis intensities of BG pulse beam corresponding to the terms $I_0$ and $I_1$ in (25): (a) $z$ = 60 mm; (b) $z$ = 90 mm. $a_0$ = 100 μm; $w_B$ = 30 μm; $\tau$ = 10 fs; $\lambda$ = 710 nm.

Variation of the propagation velocity with distance is due to the interference of propagating modes, so the superluminal and subliminal behaviors can be observed.

Note that the arrival time and velocity of the pulse beam depend on the measurement method (size and position of the receiver's aperture). The central and periphery parts of the beam propagate at different velocities, therefore, depending on the size of the aperture of the photodetector and its position in space, the velocities will be different. When only a part of the beam cross section is registered (for example, at $\rho = 0$), both the phase and group velocities of the modes affect the pulse velocity. Since the phase velocities of the modes exceed the velocity of light $c$, a superluminal propagation of the pulse can be observed near the source, i.e. the pulse propagates over small distances faster than the speed of light in vacuum. However, the average velocity determined by arrival time of the pulse at $z > l_d$ is less than $c$, even though the pulse beam experiences superluminal behaviour at some part of the propagation distance.

Given the entire beam cross section (total pulse power) have been recorded, averaging over the beam cross section occurs. In this case, the pulse velocity is determined only by the group velocities of the modes, i.e. the measured pulse velocity will be less than the speed of light $c$.

## 4. Discussion

Note that the energy velocity is always less than *c*, since it is determined only by the group velocities of the propagating modes.
Superluminal behavior can also be caused by the proximity of the receiving antenna to the emitter, i.e. when the distance between the emitter and receiver $d < c\tau$. It follows from the simulation that the center of gravity of the pulsed beam exhibits significant superluminal behavior throughout the propagation. This effect was apparently observed experimentally in [18], where a noticeable superluminality for $z < 1$ m was detected. In Fig. 7 the delay times of a microwave pulse with duration of 2 *ns* depending on the distance *z* are presented. The angular frequency of the carrier is 8.6 GHz ($\lambda$ = 3.5 cm). It can be seen that the time delay determined by the time of arrival of the pulse amplitude is almost the same as the one determined by the speed of light *c* for all distances. However, the time delay of the pulse center of gravity differs significantly from the one that is defined by the arrival time of the pulse amplitude for the distances of 1.4 *m*. The instantaneous (punctual) velocity of the pulse center of gravity in this region is higher than *c*, i.e. the superluminal behavior can be observed. This result is in good agreement with the data of [18], where superluminal behavior during microwave propagation was observed experimentally. Note that there is no such superluminality in the propagation velocity of the pulse amplitude (Fig. 7).

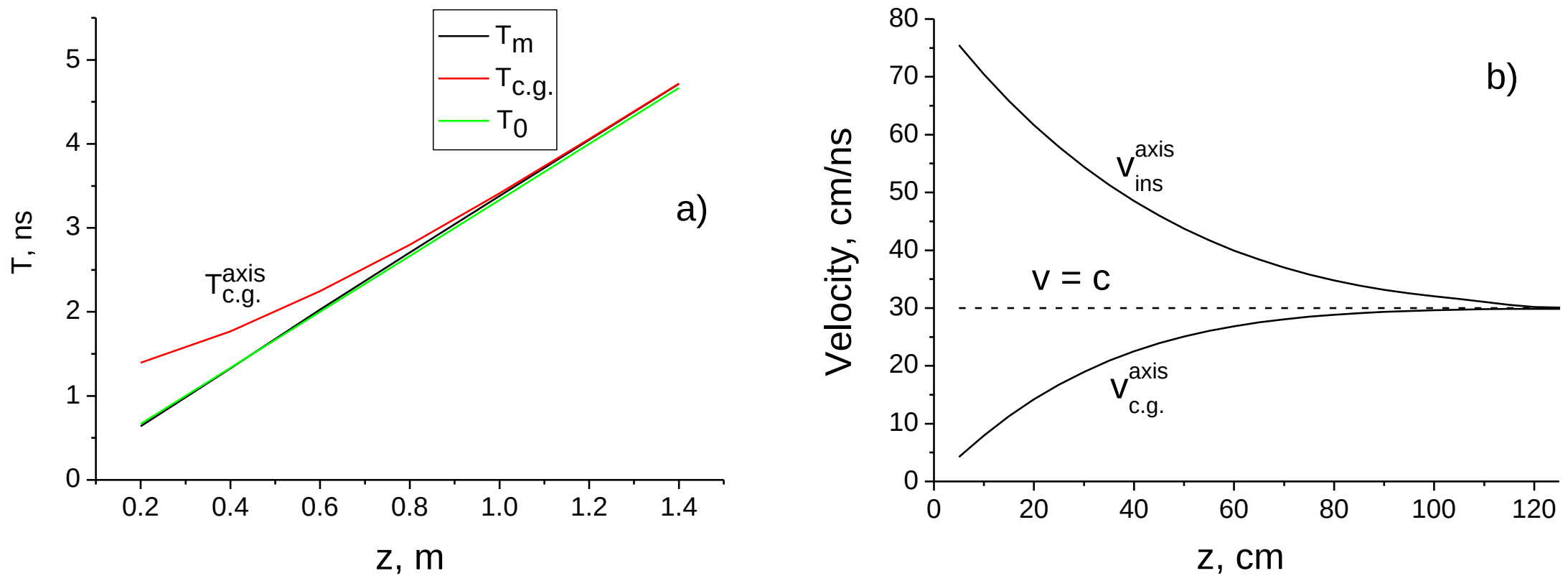

Fig. 7. (a) Arrival times $T_{c.g.}$ (red – pulse center of gravity), $T_m$ (black – pulse maximum), $T_0 = z/c$ (green – light speed *c*) of the microwave pulse (a) and velocities (b) as a function of distance *z*. $\tau$ = 2 ns; $\lambda$ = 3.5 cm.

The considered effects become significant for strongly focused pulsed beams. It is known that in this case nonparaxial effects become noticeable [19]. In this case the influence of the longitudinal field component $E_z$ and the spin-orbit interaction should be taken into account [20 - 22].
Slowing down of light propagation velocity in optical waveguides is well known and can be explained by the delays of modes. In [23] it is shown that such a delay also depends on the magnitude of the orbital angular momentum (helicity of the wave front) and the spin (polarization) of the propagating beam.
The obtained results do not contradict the experimental data obtained for femtosecond, picosecond, and nanosecond time-domain pulses in the visible and THz regions.
A number of effects have been discovered where light travels over small distances faster than the speed of light in vacuum. In this regard, the question arises whether it is possible to exchange information superluminally [24, 25]. As it was emphasized in [25], the information content of the signal tends to reduce during superluminal propagation due to noise and for this reason superluminal information exchange is impossible. Here, it is shown that the superluminal signal can be obtained only for the axial part of the incident pulsed beam. Only an axial point on the cross-section of the pulsed beam can propagate for short distances at the velocity $v > c$. The entire pulse beam always propagates with $v < c$. Hence, the information (for example, an image encoded with spatial beams) carried by the entire pulse beam can be transmitted only subluminally. Indeed, a single pixel of the image does not transmit information about the entire image, or the image information cannot be decoded from a point (pixel) on the receiver plane. Despite the fact that individual pixels can be transmitted superluminally, they do not contain all of the original information. If the receiver detects a total incident pulse beam, the superluminal signal transmission does not occur.

It was experimentally established that single photons travel at the group velocity [26]. In [13] it was shown that transverse structuring of the photon results in a decrease of the group velocity along the axis propagation. It has been now shown that the on-axis part of the pulsed beam propagates at variable speed, exhibiting both superluminal and subluminal behaviors. The effect comes from a rigorous calculation of the impulse evolution.

Usually, the terms “group” and “phase” velocities are used to describe the pulse velocity. However, these velocities are well defined only for plane waves. In the case of spatially structured pulsed beams, the term “impulse velocity” is more appropriate (see also [27]). Indeed, the impulse velocity incorporates the group and phase velocities of a set of spatial mode components.

The superluminal effect is due to interference between different modes (second term in (8)). The interference terms disappear due to the orthogonality of the modal functions if integration across the entire beam cross-section is taken. Therefore, the superluminal effect can be observed if only the pulsed beam electric field is measured at one point in the transverse plane (for example, $\rho = 0$). There is no superluminal effect if the entire cross section of the pulse beam is recorded.

## 5. Conclusions

Thus, the theoretical analysis of the nonparaxial propagation of localized wave packets in free space on the basis of the wave approach is carried out. The modal approach is shown to provide a clear physical insight in the subluminal and superluminal behaviors of pulsed beams in free space, which arise from the interference of modes and depend on the parameters of the pulsed beam, the measurement system and the range of propagation distance.

The importance of the measurement method in the observation of the superluminal effect is emphasized. Usually the spatial dimensions of receivers are smaller than the transverse beam sizes. This indicates that the interference terms in (8) will contribute to the velocity of the pulse beam at the measurements. It is shown that superluminal propagation occurs due to interference between spatial modes, and its observation is possible if only a part of the beam cross-section is recorded.

It is shown that the pulse beam propagates in free space with variable velocity along the axis. This indicates that the average propagation velocity of a pulsed beam with specified initial parameters (pulse duration, beam radius, OAM, and frequency spectrum) depends on the distance between the source and the receiver. The change in propagation velocity along the axis with distance is due to the interference of propagating modes, so superluminal and subliminal behaviors can be observed along the propagation axis. It is shown that the on-axis velocity varies with distance approaching the fundamental value of $c$ at large distances. The velocity of the total pulsed beam is always less than the velocity of the plane wave $c$ and remains constant during propagation.

Although the local on-axis velocity of the pulse may be higher than $c$, the average velocity, which is estimated from the arrival time of the pulse at $z > l_d$, is less than $c$. The obtained results do not contradict the experimental data on the observation of superluminal and subluminal effects.

The slowing down of the speed of propagation with a decrease in the spatial dimension of the incident pulsed beam is shown. This can be used, for example, in all-optical switching using slow light [28].

In conclusion, the effect of inconstancy of the pulse beam propagation velocity with a distance arises due to the interference of propagating modes, so it is possible to observe superluminal and subliminal behaviors. It is shown that the observation of superluminal propagation is possible if only a part of the beam cross-section is recorded. Although the decrease in the speed of light for paraxial beams is hardly noticeable, it should be taken into account when accurate distance determination is required. These results are particularly important in applications such as time-of-flight measurements, radio and satellite communications, free-space optical communication, as well as in quantum information and gravitational wave experiments.